\documentstyle[sprocl,epsfig]{article}

\def\Journal#1#2#3#4{{#1} {\bf #2}, #3 (#4)}

\def\NPB{{\em Nucl. Phys.} B}
\def\PLB{{\em Phys. Lett.}  B}

\def\PRD{{\em Phys. Rev.} D}

\newcommand{\bm}[1]{\mbox{\boldmath $#1$}}
\newcommand{\beq}{\begin{equation}}
\newcommand{\eeq}{\end{equation}}
\newcommand{\beqq}{\begin{eqnarray}}
\newcommand{\eeqq}{\end{eqnarray}}
\newcommand{\Lup}{\Lambda^\uparrow}

\begin{document}

\title{
LONGITUDINAL \bm{\Lambda} POLARIZATION IN POLARIZED SEMI-INCLUSIVE  
DIS\footnote{
Talk delivered by U.~D'Alesio at the IX International
Workshop on Deep Inelastic Scattering (DIS2001), Bologna, 
27 April - 1 May, 2001.}}

\author{M. ANSELMINO$^\flat$, M. BOGLIONE$^\dag$, 
U. D'ALESIO$^{\dag\dag}$, F. MURGIA$^{\dag\dag}$}

\address{\vspace{4pt}
$^\flat$ Dipartimento di Fisica Teorica, Universit\`a di
Torino and \\
INFN, Sezione di Torino, Via P.~Giuria 1, I-10125 Torino, Italy\\
\vspace*{4pt}
$^\dag$ Department of Physics, University of Durham, Sciences
Laboratories, \\
South Road, Durham DH11 3LE, United Kingdom\\
\vspace*{4pt}
$^{\dag\dag}$ Dipartimento di Fisica, Universit\`a di Cagliari and 
INFN, Sezione di Cagliari, \\
C.P. 170, I-09042 Monserrato (CA), Italy
}

\maketitle

\abstracts{
We calculate, within pQCD parton model at leading orders, the 
expression of the longitudinal polarization of  
$\Lambda$ baryons produced in polarized semi-inclusive DIS, 
including weak interaction effects.
We present some numerical estimates in few cases for which data are or 
will soon be available and discuss how to gather new information on 
polarized fragmentation functions.}

\section{Introduction}

The polarization of spin 1/2 baryons inclusively produced in polarized
Deep Inelastic Scattering processes may be useful, if measurable, 
to obtain new information on polarized distribution and fragmentation 
functions. 

Because of their parity violating weak decay and their  
self-analysing polarization, $\Lambda$ hyperons are an ideal
laboratory for this study. 
Moreover, since in a non-relativistic quark model 
the spin of a $\Lambda$ is carried by the strange
quark, any signal of an up (or down) quark spin contribution would
reveal a novel hadron spin structure.

From the experimental side, 
NOMAD collaboration have recently published new results
\cite{nom} on the $\Lambda$ polarization in $\nu_\mu$ charged current 
interactions; more data might soon be available from high energy neutral
current processes at HERA, due to electro-weak interference effects.
It is then timely to perform a systematical and comprehensive 
study of weak interaction contributions to the production and the polarization
of $\Lambda$ 
baryons in as many as possible DIS processes.
We stress that such contributions are an important source of new 
information, due to the natural neutrino polarization and to the selected 
couplings of $W$'s to pure helicity states. 

In our calculations we take into account leading twist factorization theorem, 
Standard Model elementary interactions at lowest perturbative order and 
LO QCD evolution only. 

\section {Charged current processes, \bm{\nu p \to \ell \Lup X}}

For these processes, there exist two possible elementary 
contributions, corresponding to the interactions:
$\nu  d (s)  \to \ell^- u$ and 
$\nu \bar u  \to \ell^- \bar d(\bar s) $. 
Neglecting quark masses one finds that there is only one non-zero helicity
amplitude for each of them,  
corresponding to the following elementary cross-sections:
$d\hat\sigma^{\nu q}_{--}$ and 
$d\hat\sigma^{\nu \bar q}_{-+}$. 
For longitudinally ($\pm$ helicity) polarized protons
we have \cite{paper} 
\beq
P_{[\nu,\ell]} ^{(\pm)} (B;x,y,z) = 
-\frac{[d_\mp + R\,s_\mp]\, \Delta D_{B/u} - 
(1-y)^2 \,\bar u_\pm \,[\Delta D_{B/\bar d} + R\, \Delta D_{B/\bar s}]}
{[d_\mp + R\,s_\mp] \, D_{B/u} +
(1-y)^2 \,\bar u_\pm \, [D_{B/\bar d} + R\, D_{B/\bar s}]} 
\label{Pnu+-}
\eeq
and 
\beq
P_{[\bar \nu,\ell]} ^{(\pm)} (B;x,y,z) = 
\frac{[\bar d_\pm + R\, \bar s_\pm]\, \Delta D_{B/\bar u} - 
(1-y)^2 \,u_\mp \, [\Delta D_{B/d} + R\, \Delta D_{B/ s}]}
{[\bar d_\pm + R\, \bar s_\pm] \, D_{B/\bar u} +
(1-y)^2 \, u_\mp \, [D_{B/ d} + R\, D_{B/s}]} \,,
\label{Pantinu+-}
\eeq
where $(D_{B/q})$ 
$\Delta D_{B/q}$ are the (un)polarized fragmentation functions (FF) and 
$R\, \equiv \tan ^2 \theta_C \simeq 0.056$. 
In the case of unpolarized protons one replaces
the polarized parton distributions $q_+$ and $q_-$ with $q/2$.

When a $\Lambda$ baryon is produced (rather than a $\bar\Lambda$), 
we can neglect terms which contain 
both $\bar q$ distributions 
and $\bar q$ fragmentations 
and we simply have:  
\beqq
P_{[\nu,\ell]} ^{(\pm)} (\Lambda;z) & \simeq & 
P_{[\nu,\ell]} ^{(0)} (\Lambda;z) \; \simeq \;  
-\frac{\Delta D _{\Lambda/u}}{D_{\Lambda/u}}\,, \label{PnuL}\\
P_{[\bar \nu,\ell]} ^{(\pm)}  (\Lambda;z) & \simeq & 
\, P_{[\bar \nu,\ell]} ^{(0)} (\Lambda;z) \;\simeq \;
- \frac{\Delta D _{\Lambda/d} + R\,\,\Delta D _{\Lambda/s}}{D_{\Lambda/d} 
+ R\,\, D _{\Lambda/s}}\,, \label{PantinuL}
\eeqq
and the polarizations, up to QCD evolution effects, become functions of 
the variable $z$ only.

\begin{figure}[t]
\begin{center}
\hspace*{8pt}
\epsfig{file=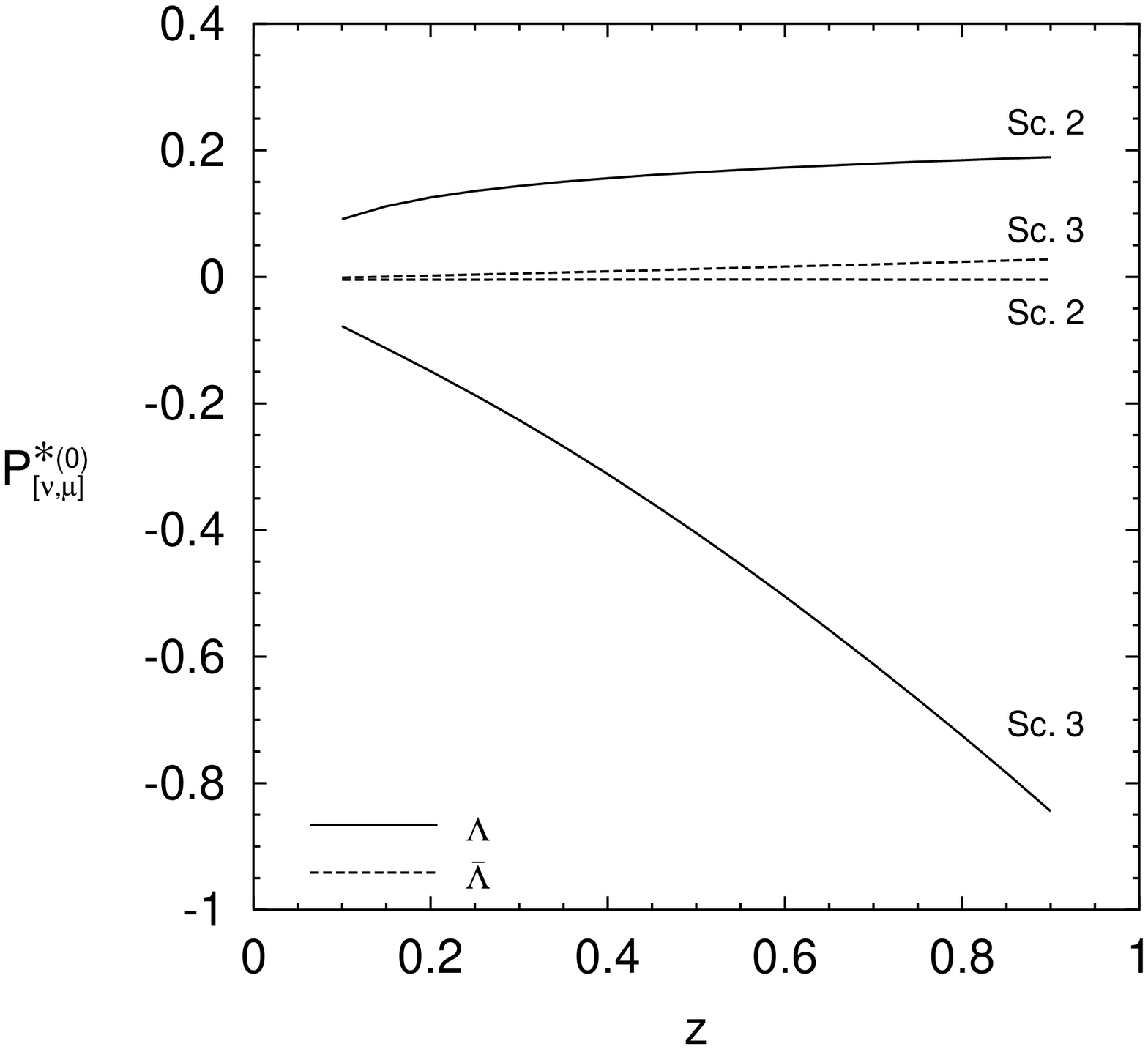,angle=-90,width=5.3cm} \hspace*{5pt} 
\epsfig{file=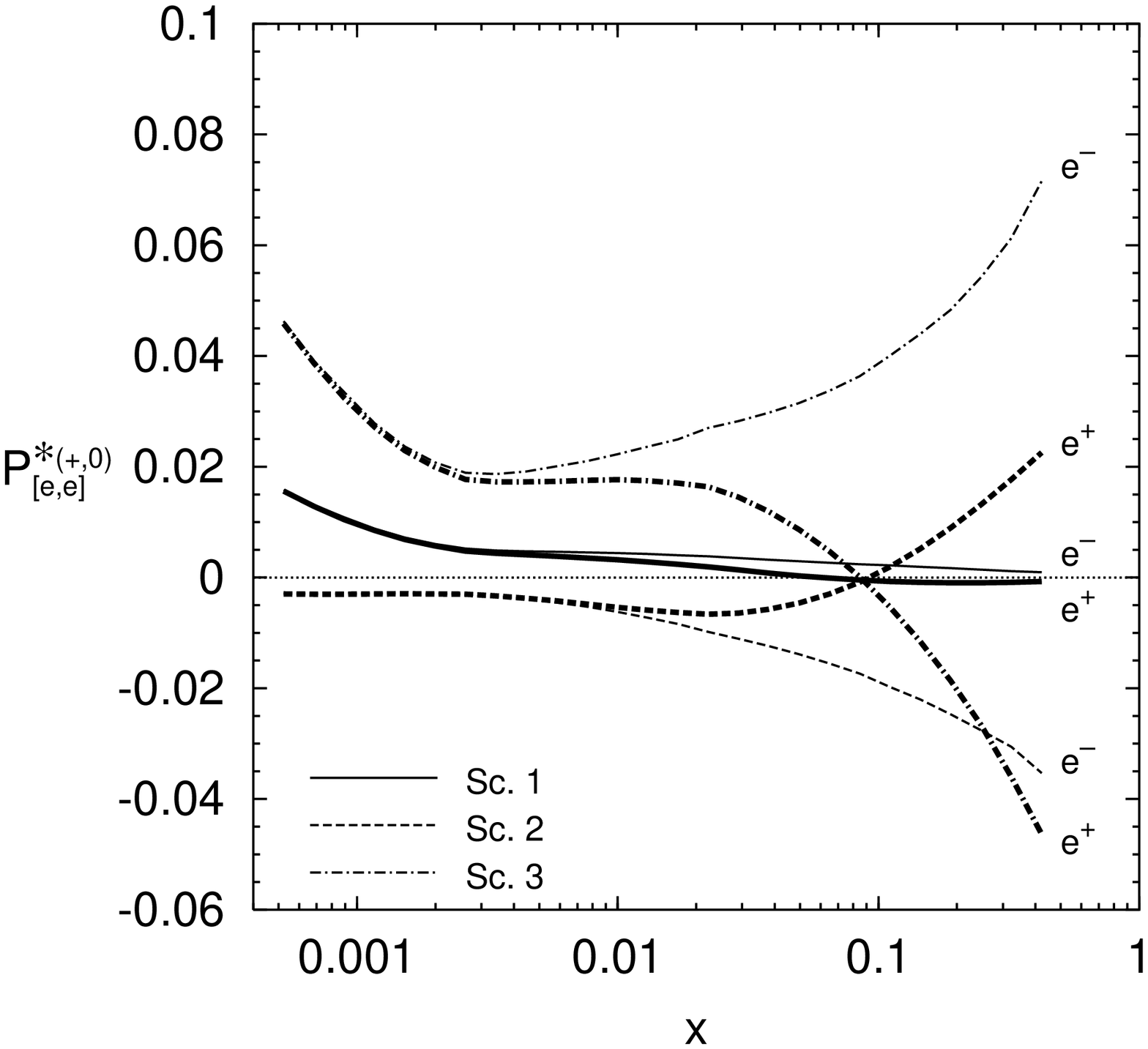,angle=-90,width=5.3cm} 
\end{center}
\caption[a1]{\label{figs}
Left: $P^{\ast\,(0)}_{[\nu,\mu]}$ as a function of $z$,
with a kinematical setup typical of NOMAD experiment at CERN. 
Results with scenario 1 are almost negligible. 
Since $\langle{x}\rangle$ is large for this kinematical
configuration, we expect
$T=d\sigma^{\bar\Lambda}/d\sigma^\Lambda \ll 1$ and, as a consequence,
$P^{(0)}_{[\nu,\mu]}(\Lambda)\simeq
P^{*\,(0)}_{[\nu,\mu]}(\Lambda)$, while
$P^{(0)}_{[\nu,\mu]}(\bar\Lambda) \gg
P^{*\,(0)}_{[\nu,\mu]}(\bar\Lambda)$. 
Right: 
$P^{*(+,0)}_{[e,e]}$ for $\Lambda$ hyperons, as a function of $x$.
The kinematical setup is typical of HERA experiments at DESY. 
Since $T \ll 1$ at large
$x$ and becomes comparable to unity at very low $x$, we expect, 
correspondingly, $P_{[e,e]}(\Lambda)\simeq P^{*}_{[e,e]}(\Lambda)$ and 
$P_{[e,e]}(\Lambda, \bar\Lambda) \simeq 2 \, 
P_{[e,e]}(\Lambda,\bar\Lambda)$.
}
\end{figure}

Eqs.~(\ref{PnuL}) and (\ref{PantinuL}) relate the values of the 
longitudinal polarization $P(\Lambda)$ to a quantity with a clear physical 
meaning, {\it i.e.} the ratio $C_q \equiv 
\Delta D_{\Lambda/q}/D_{\Lambda/q}$; 
this happens with weak charged current interactions -- while it cannot happen 
in purely electromagnetic DIS \cite{noi2} -- due to the selection of the quark 
helicity and flavour in the coupling with neutrinos. 
In Fig.~\ref{figs} (left) some predictions for 
$P(\Lambda)$ and $P(\bar\Lambda)$ 
are shown (see also Sec.~\ref{result}).

Notice that a comparison of $P(\Lambda)$ for $\nu$ and $\bar\nu$ beams  
might give information on the ratios $C_q$;  
for example, the same value of $C_q$ for all flavours would result in 
$P_{[\nu,\ell]}(\Lambda) = P_{[\bar\nu,\ell]}(\Lambda)$.
On the other hand, largely different values of 
$P_{[\nu,\ell]}$ and $P_{[\bar\nu,\ell]}$ would certainly indicate a strong 
$SU(3)$ symmetry breaking in the fragmentation functions.

\section {Neutral current lepton processes, \bm{\ell p \to \ell \Lup X}}

In this case there are four non-zero independent helicity 
amplitudes for the elementary processes $\ell q\to \ell q$; 
both weak and electromagnetic interactions are 
to be included at the amplitude level. 
The complete (long) formulae~\cite{paper} are not shown here; 
some interesting results are displayed in
Fig.~\ref{figs} (right), see also Sec.~\ref{result}, 
where we plot the $\Lambda$ polarization 
vs.~$x$ (averaged over $z$) for longitudinal polarized leptons and
unpolarized protons. 
These results test the dynamics of the partonic 
process and in particular the contribution of electro-weak interferences,  
in a neat and unusual way. The differences between positron and 
electron beams are entirely due to electro-weak effects; 
this is well visible at large $x$, 
where the curves for $e^+$ and $e^-$ differ sizeably. 
Moreover for unpolarized leptons and protons   
the longitudinal $\Lambda$ polarization  is 
non-zero~\cite{paper} only due to parity violating weak contributions.

\section {Numerical estimates}
\label{result}

The polarization values depend on the known Standard Model dynamics,
on the rather well known partonic distributions, 
and on the quark fragmentation functions. 
The latter are not so well known and a choice 
must be made in order to give numerical estimates or in order to be able 
to interpret the measured values in favour of a particular set.

Unpolarized $\Lambda$ FF are determined by fitting 
$e^+e^- \to (\Lambda + \bar \Lambda) + X$  data, sensitive 
only to singlet combinations, like $D_{\Lambda/q} + D_{\Lambda/\bar q}
\equiv D_{(\Lambda + \bar\Lambda)/q}$. 
Polarized $\Lambda$ FF are obtained by fitting 
the scarce data on $\Lambda$ polarization at LEP, sensitive only to 
non-singlet combinations like 
$\Delta D_{\Lambda/q} - \Delta D_{\Lambda/\bar q} \equiv  
\Delta D^{val}_{\Lambda/q}$. In both cases flavour separation has
to rely on models.  

We adopt three  sets of FF, denoted as scenarios 1, 2 
and 3, and derived~\cite{vog} from fits to $e^+e^-$ data for which  
$\Delta D_{\Lambda/u(d)} \simeq N_u\,\Delta D_{\Lambda/s}$   
($D_{\Lambda/u(d)} = D_{\Lambda/s}$). 
The three scenarios differ for the relative contributions of the 
strange quark polarization to $\Lambda$ polarization: $N_u = 0$, $N_u = -0.2$
and $N_u = 1$ for scenarios 1, 2 and 3 respectively and are 
well representative of possible spin dependences.

Equipped with unpolarized FF into 
$\Lambda + \bar\Lambda$ and with separate polarized FF 
into $\Lambda$ and $\bar\Lambda$,  
we can define the following computable quantities:
\beq
P^*(\Lambda) \equiv \frac{d\sigma^{\Lambda_+} - d\sigma^{\Lambda_-}}
{d\sigma^{\Lambda + \bar\Lambda}} = 
\frac{P(\Lambda)}{1 + T} \,,\label{pstarl}
\quad
P^*(\bar\Lambda) \equiv \frac{d\sigma^{\bar\Lambda_+} - 
d\sigma^{\bar\Lambda_-}}
{d\sigma^{\Lambda + \bar\Lambda}} = 
\frac{T\,P(\bar\Lambda)}{1 + T} \,,\label{pstarbarl}
\eeq
where $T = d\sigma^{\bar\Lambda}/d\sigma^{\Lambda}$. 
Eqs.~(\ref{pstarl}) allow to compute the values of
$P(\Lambda)$ and $P(\bar\Lambda)$ provided one can compute or measure
the ratio $T$. 
Notice that $P$ is always larger in magnitude than $P^*$. 

We conclude by pointing out how
this comprehensive study of the polarization of $\Lambda$'s and
$\bar\Lambda$'s  can help to gather new
information about polarized FF  
and to test fundamental features of electro-weak elementary interactions.

\vspace*{5pt}

U.D. and F.M. thank COFINANZIAMENTO MURST-PRIN for partial support.

\end{document}